\documentclass[12pt]{amsart}                                              
\usepackage{amssymb}
\usepackage{float,url,verbatim}
\usepackage[svgnames]{xcolor}
\usepackage{hyperref}
\usepackage[leftmargin=2.5em,rightmargin=2.5em]{quoting}
\usepackage[all,cmtip]{xy}
\newtheorem{theorem}{Theorem}[section]

\theoremstyle{definition}
\newtheorem{definition}[theorem]{Definition}

\newtheorem{examples}[theorem]{Examples}

\theoremstyle{remark}

\newenvironment{qna}
 {\bigskip\noindent
  \hrulefill\hfill\hfill\mbox{}\\[-2em]\par}
 {\\[-2em]\par\noindent\hrulefill\hfill\mbox{}
  \bigskip}
\newenvironment{qna-}
 {\bigskip\noindent
  \hrulefill\hfill\hfill\mbox{}\\[-2em]\par}
 {}
\newcommand{\Qn}{\medskip\noindent\textbf{Q:\ }}
\newcommand{\Ar}{\medskip\noindent\textbf{A:\ }}

\renewcommand{\1}{^{-1}}
\newcommand{\A}{\ensuremath{\mathcal A}}
\newcommand{\fs}{\ensuremath{\texttt{FinSet+Iso}}}
\newcommand{\N}{\mathbb N}
\newcommand\xqed[1]{%
  \leavevmode\unskip\penalty9999 \hbox{}\nobreak\hfill
  \quad\hbox{#1}}
\newcommand\textqed{\xqed{$\triangleleft$}}

\title{Who needs category theory?}
\author{Andreas Blass}
\address{Mathematics Department\\
University of Michigan\\
Ann Arbor, MI 48109--1043, U.S.A.}
\email{ablass@umich.edu}
\author{Yuri Gurevich}
\address{Computer Science and Engineering\\
University of Michigan\\
Ann Arbor, MI  48109-2121, U.S.A}
\email{gurevich@umich.edu}

\begin{document}
\maketitle
\thispagestyle{empty}
\begin{abstract}
In mathematical applications, category theory remains a contentious issue, with enthusiastic fans and a skeptical majority. In a muted form this split applies to the authors of this note. When we learned that the only mathematically sound foundation of topological quantum computing in the literature is based on category theory, the skeptical author suggested to ``decategorize" the foundation. But we discovered, to our surprise, that category theory (or something like it) is necessary for the purpose, for computational reasons. The goal of this note is to give a high-level explanation of that necessity, which avoids details and which suggests that the case of topological quantum computing is far from unique.
\end{abstract}

\section{Introduction}
\label{sec:intro}

Category theory is indispensable in some parts of mathematics, e.g.\ in algebraic geometry, homological algebra, algebraic topology. Yet, even among mathematicians the attitude toward category theory varies greatly, as witnessed by the following joke of John Baez \cite{Baez}.

\medskip
\begin{quoting}\footnotesize\sf
	\noindent
	I hope most mathematicians continue to fear and despise category theory,\\ so I can continue to maintain a certain advantage over them.
\end{quoting}

\medskip
The chasm between the fans of category theory and the silent majority is even more pronounced in computer science where the fans tend to be super-enthusiastic while the majority is indifferent.

In a muted form this split applies to the authors of this note. As we mentioned in \cite[\S1]{G225}, ``The first author of this paper has long been a fan of category theory; even as a graduate student, he was described by one of his professors as `functorized'. The second author has been far more skeptical about the value of category theory in computer science."

It turns out, however, that the only mathematically sound foundation of topological quantum computing in the literature is based on category theory; see \cite{G225,PP,Wang} for example. Why? Is this just an accident of history or there is more to it?
We have been debating this question for a while, and now we agree that something like category theory is necessary for the purpose.

Categories were introduced by Samuel Eilenberg and Saunders Mac Lane as an auxiliary notion in their general theory of natural equivalences. Here we argue that something like categories is needed on a more basic level. As you work with operations on structures, it may be necessary to coherently manipulate witnesses for various properties of these operations, e.g.\ isomorphisms witnessing associativity, commutativity and distributivity of the operations.
A working mathematician, to use Mac Lane's term, is well advised to be aware of the coherent witness-manipulation problem and to know that category theory or something like it can provide an appropriate framework to address the problem. Of course, the working mathematician in question may be a computer scientist or physicist.

\section{What's category theory?}
\label{sec:cats}

For those who have only a vague idea of category theory, let us say a few words about it. The experts can safely skip this section.

Categories were introduced by Samuel Eilenberg and Saunders Mac Lane as an auxiliary notion in their general theory of natural equivalences \cite{EM}. ``It is not too misleading, at least historically, to say that categories are what one must define in order to define functors, and that functors are what one must define in order to define natural transformations,'' writes Peter Freyd in the introduction to his book \cite{Freyd}.

In the rest of this section, we quickly explain the notion of natural equivalence.

\subsection{Categories and functors}

\begin{definition}\label{def:cats}
A \emph{category} comprises
\begin{enumerate}
\item a collection of \emph{objects},
\item for any objects $x,y$, a collection of \emph{arrows} $\alpha: x\to y$\\including, if $x=y$, an \emph{arrow} $1_x : x\to x$\ \ called \emph{identity},
\item a \emph{composition} $\beta\alpha:x\to z$ of arrows
$x\overset{\alpha}\longrightarrow y\overset{\beta}\longrightarrow z$
which is associative and treats the identity arrows as expected.
\end{enumerate}
An arrow with a two-sided inverse is an \emph{isomorphism}.
\textqed
\end{definition}

\begin{examples}\label{exs:1}\mbox{}
\begin{enumerate}
\item Sets, total functions, and function composition.
\item Groups, group homomorphisms, and function composition.
\end{enumerate}
\end{examples}

We will be using below a restricted version of the first of the two examples where objects are finite sets and arrows are isomorphisms, that is, one-to-one correspondences.
For brevity, this category of finite sets with isomorphisms will be denoted \fs.

\begin{definition}
Let $C,D$ be categories. A \emph{functor} $F:C\to D$ is a mapping from $C$-objects and $C$-arrows to $D$-objects and $D$-arrows respectively such that
\begin{itemize}
\item if $\alpha:x\to y$ then $F\alpha: Fx\to Fy$,
\item $F1_x= 1_{Fx}$,
\item $F(\beta\alpha) = (F\beta) (F\alpha)$.
\end{itemize}
\end{definition}

\begin{examples}\label{exs:2}
We describe two functors $F$ and $G$ from \fs\ to \fs. Let $S$ and $S'$ be arbitrary finite sets.
\begin{enumerate}
\item $F(S)$ is the set of all permutations of $S$, that is, bijections from $S$ to $S$. For any isomorphism $\alpha: S\to S'$, $F\alpha$ transforms every permutation $\pi$ of $S$ into a permutation $\alpha\pi\alpha\1$ of $S'$. That is, if $\pi$ maps $x$ to $y$ then $(F\alpha)(\pi)$ maps $\alpha x$ to $\alpha y$. It is easy to check that $F(\beta\alpha) = (F\beta)(F\alpha)$.
\item $G(S)$ is the set of linear orderings of $S$. For any isomorphism $\alpha: S\to S'$, $G\alpha$ transforms every linear ordering $<$ of $S$ into the ordering
    \[ s <' t \iff \alpha\1s < \alpha\1t \]
    of $S'$. It is easy to check that $G(\beta\alpha) = (G\beta)(G\alpha)$.
\end{enumerate}
\end{examples}

\subsection{Natural equivalences}\mbox{}\medskip

We start with a motivating example. For any finite set $S$, there are as many permutations of $S$ as linear orderings. If $n$ is the cardinality of $S$ then there are $n!$ permutations and $n!$ linear orderings. It follows that there is a bijection between the permutations and linear orderings of $S$.

If $n\ge2$, there are multiple such bijections. Yet, on the level of abstraction where you don't distinguish between elements of $S$, you cannot single out any such bijection. The reason is that linear orderings are all automorphic while permutations are not. For example, the identity permutation is preserved by all automorphisms.

On the other hand, suppose that $S$ comes furnished with a particular linear order
\[ s_1, s_2, \dots, s_n \]
where $n$ is the cardinality of $S$. Then we have a standard bijection, let us call it $\tau_0$, between the permutations and the linear orders: $\tau_0(\pi)$ is the linear order
\[ \pi(s_1), \pi(s_2), \dots, \pi(s_n). \]
This bijection is natural in that it works uniformly for any finite set $S$ furnished with a linear order.

\begin{definition}
Given two functors $F,G: C\to D$, a \emph{natural transformation} $\tau$ of $F$ to $G$ assigns to each object $x$ of $C$ an arrow $\tau x: Fx\to Gx$ of $D$ in such a way that every arrow $\alpha: x\to y$ in $C$ yields a commutative diagram
\begin{figure}[H]
\[\xymatrix@C+2pc@R+2pc
{
  Fx \ar[r]^{\tau x}
     \ar[d]_{F\alpha}
& Gx \ar[d]^{G\alpha}
\\
  Fy \ar[r]^{\tau y}
& Gy
}\]
%\caption{11.1}
\end{figure}
\noindent
Further, $\tau$ is a \emph{natural equivalence} if every $\tau x$ is an isomorphism. \textqed
\end{definition}

Coming back to the motivating example, let $C$ and $D$ be the category \fs. The functors $F,G$ of Examples~\ref{exs:2} are not naturally equivalent, for the reason mentioned above. Indeed, suppose toward contradiction that $\tau$ is a natural equivalence from $F$ to $G$. Let $x$ and $y$ be the same set $S=\{a,b\}$, so that there are two permutations of $S$,
$\pi_1 = \left(\begin{smallmatrix}a&b\\a&b\end{smallmatrix}\right)$
\and $\pi_2 =    \left(\begin{smallmatrix}a&b\\b&a\end{smallmatrix}\right)$, and two linear orderings, $a<_1b$ and $b<_2a$. Let $\alpha$ transpose $a$ and $b$, so that $\alpha\1=\alpha$,
$(F\alpha)(\pi_i) = \alpha\pi\alpha = \pi_i$ and $(G\alpha)(<_i) = <_{3-i}$. If $\tau\pi_i =<_i$ then
\begin{align*}
& (G\alpha)(\tau x)(\pi_1) = (G\alpha)(<_1) = <_2 \\
& (\tau y)(F\alpha)(\pi_1) = (\tau y)(\pi_1) = <_1
\end{align*}
If $\tau\pi_i =<_{3-i}$ then
\begin{align*}
& (G\alpha)(\tau x)(\pi_1) = (G\alpha)(<_2) = <_1 \\
& (\tau y)(F\alpha)(\pi_1) = (\tau y)(\pi_1) = <_2
\end{align*}
In either case, the diagram above does not commute.

On the other hand, suppose that the domains of $F$ and $G$ are modified so that, for any object $x$, the finite set $x$ is furnished with a fixed linear order and, for any two objects, $x$ and $y$, the arrows from $x$ to $y$ respect the fixed orders, so that, in fact, there is a unique arrow from $x$ to $y$. Then
the functors $F,G$ become naturally equivalent. The standard bijection $\tau_0$, described above, makes the diagram commute. (In the case where $x$ and $y$ are the set $\{a,b\}$ with the same linear order, the transposition is not a legitimate arrow; the only legitimate arrow is the identity.)

\section{A dialog on category theory\\ and its applications in computing}
\label{sec:dialog}

\noindent\textbf{Q\footnotemark:\ }
The formalization of the intuitive notion of natural equivalence is impressive.
Can you explain other advantages of category theory to me?
Unfortunately I don’t know more category theory than you have just taught me here.

\footnotetext{Quisani, a former student of the second author}

\medskip\noindent\textbf{A\footnotemark:\ }
Let us illustrate one advantage of category theory. It is related to the notion of isomorphism. You worked with isomorphisms.

\footnotetext{The authors, speaking one at a time}

\Qn Sure. I know the notion from universal algebra in general and group theory in particular. To me, an isomorphism is a homomorphism that happens to be bijective. Judging by Examples~\ref{exs:1}, the notion of isomorphism of Definition~\ref{def:cats} is rather similar. I am thinking of arrows as homomorphisms. The two notions of isomorphism may be equivalent.

\Ar Actually, the categorical notion of isomorphism is more general. Consider the category of partially ordered sets (in short, posets) with monotone maps. Let $A, B$ be two-element posets where the two elements are incomparable in $A$ but ordered in $B$. Any bijection from $A$ to $B$ is a monotone map but categorically --- and intuitively! --- the two posets are not isomorphic.

Another category where not all bijective arrows are isomorphisms is the category of topological spaces with continuous functions. Here, isomorphisms are exactly homeomorphisms, but continuous bijections form a larger class. For example, let $A,B$ be two-element topological spaces where every subset of $A$ is open but, in $B$, only $\emptyset$ and $B$ itself are open. Any bijection from $A$ to $B$ is continuous but not a homeomorphism. For a more interesting example, let $A$ be the half-open interval $[0,1)$ of the real line and $B$ be the unit circle in the complex plane. The bijection $r\mapsto e^{2\pi i r}$ is continuous but its inverse is discontinuous at $1$.

\Qn Fascinating. Anything else?

\Ar What do these mathematical constructions
\begin{itemize}
\item free groups,
\item tensor algebra,
\item universal enveloping algebras,
\item abelianizations of groups, and
\item Stone-\v{C}ech compactifications
\end{itemize}
have in common?

\Qn I don't know. They come from different parts of mathematics and look disparate to me. Certainly, the free group construction and group abelianization are quite different. There is something universal about each of the constructions, but I don't see more than that. Oh, wait. I guess that every one of these constructions is a functor.

\Ar You are right, all these constructions can be viewed as functors. But the amazing part is that these are examples of a single categorical construction. All these functors are left adjoints of forgetful functors. Unfortunately, we haven't covered forgetful functors or adjoints here.

\Qn The unifying power of category theory seems awesome. That's got to be useful in many areas of mathematics, I guess. What about computing?

\Ar The usefulness of category theory in computing is less obvious. Until recently when we started to work on topological quantum computing (TQC for short), one of us had been skeptical.

\Qn Why?

\Ar Because of the distance of this very abstract theory from computing and because of the peril of potential (and in some cases actual) over-abstraction. There is also the hammer-and-nail phenomenon: ``For a person with a hammer, everything looks like a nail.''

\Qn You don't mean that category theory itself is an over-abstraction.

\Ar No, we don't. As Seneca the Younger said in the first century, ``gladius neminem occidit: occidentis telum est,'' that is ``a sword kills nobody; it is a tool of the killer.''

\Qn Give me a relevant example of that hammer-and-nail phenomenon.

\Ar Here is a true life example, but allow us to omit the reference. A computation can be seen as a category where objects are states and morphisms are state transitions. If you take this point of view, then you might want computation transformers to be functorial, which narrows unreasonably your library of computation transformers. For example, you lose compilers.

\Qn Why isn't a compiler functorial?

\Ar Typically, the target language is at a lower abstraction level and uses different data structures. Some higher-level steps may have no meaning at the lower level. Besides, think of compiler optimization.

\Qn How did topological quantum computing influence the skepticism? And what is topological quantum computing?

\Ar
Topological quantum computation employs two-dimensional quasiparticles called anyons \cite{FKLW,Kitaev}. What is relevant for our purposes here is that the generally accepted mathematical basis for the theory of anyons is the framework of modular tensor categories. That framework, as presented in \cite{Wang} or \cite{PP} or \cite{G225} involves a substantial amount of category theory and is, as a result, considered rather difficult to understand.

Why is the only mathematically sound theory of anyons in the literature based on category theory? The skeptic among us suspected that this is just an accident of history, another nail for the categorical hammer. Hence the idea to ``decategorize" the theory of anyons.

As we worked on the decategorization project, we realized that, surprisingly, category theory --- or something like category theory --- is necessary for the theory of anyon computations.

\Qn Can you explain to me, who knows nothing about anyons, why category theory is necessary for the purpose? Is the reason specific to the anyon theory?

\Ar The reason seems to us more generic and not at all specific to the anyon theory, but at this point we do not have other natural examples where category theory is necessary for the same reason.

In the next section, we will try to illustrate the reason behind the necessity of category theory (or something like it) for the theory of anyons.

\section{Coherent witness manipulation}
\label{sec:main}

We illustrate why (something like) category theory is needed in topological quantum computing.

\subsection{Algebra of structures}\mbox{}

Consider an algebra $\A$ of structures (that is, elements of \A\ are structures) together with operations of addition $+$ and multiplication $\ast$ where, up to isomorphism,
\begin{itemize}
\item both operations are commutative and associative,
\item both operations have their respective neutral elements $\boldsymbol 0$ and $\boldsymbol 1$, and
\item multiplication distributes over addition.
\end{itemize}

The following example is a simplified version of the algebra used in topological quantum computing.
\begin{itemize}
\item The structures in $\A$ are finite-dimensional vector spaces, over the field of complex numbers, each furnished with a fixed basis.
\item The vector space $A+B$ is the direct sum, also known as the direct product, of vector spaces $A$ and $B$ furnished with the disjoint union (of a particular form\footnotemark) of the fixed bases of $A$ and $B$.
\item The product $A\ast B$ is the tensor product of the vector spaces $A$ and $B$ furnished with the cartesian product of the fixed bases of $A$ and $B$.
\end{itemize}
\footnotetext{In the standard construction of $A+B$, the base set is the set of ordered pairs $(a,b)$ where $a\in A$ and $b\in B$. With that convention, the standard basis for $A+B$ consists of vectors $(a,0)$ for $a$ in the standard basis of $A$ and $(0,b)$ for $b$ in the standard basis of $B$.}

In the case of topological quantum computing, the structures are more sophisticated --- involving e.g.\ tuples of Hilbert spaces, duality, ribbon structures --- but this is not important. For the purposes of this note, we may and will pretend that the example above is the one used in topological quantum computing.

So far, we are within the realm of universal algebra.
But we need to go beyond universal algebra.

\subsection{Witness isomorphisms}

Suppose that algebra $\A$ is equipped with standard isomorphisms
\begin{align*}
&\texttt{associative +\qquad}
&&\alpha^{+}_{A,B,C}:&&(A+ B)+ C&&\to&& A+(B+ C)\\
&\texttt{commutative +}
&&\gamma^+_{A,B}:&& A+ B&&\to&& B+ A\\
&\texttt{associative }\ast
&&\alpha^{\ast}_{A,B,C}:&& (A\ast B)\ast C&&\to&& A\ast(B\ast C)\\
&\texttt{commutative }\ast
&&\gamma^\ast_{A,B}:&& A\ast B&&\to&& B\ast A\\
&\texttt{distributive}
&&\delta_{A,B,C}:&& A\ast(B+ C)&&\to&& (A\ast B)+(A\ast C)
\end{align*}
working properly with $\boldsymbol 0$ and $\boldsymbol 1$.

It is convenient to think of an isomorphism $\xi: A\to B$ as a \emph{witness} that $A,B$ are isomorphic. Accordingly, the standard isomorphisms above are \emph{standard witnesses}.

There are numerous requirements that we have to impose on the
standard isomorphisms. In particular, it is required that $\gamma^{+}_{A,B} = (\gamma^{+}_{B,A})\1$. This property is called \emph{symmetry}. Thus the additive structure of $\A$ is symmetric.

A relevant peculiarity of topological quantum computing is that the multiplicative structure is not symmetric. It is not required that $\gamma^{\ast}_{A,B}$ coincides with $(\gamma^{\ast}_{B,A})\1$. It is convenient to think about this topologically: as $A\ast B$ is transformed into $B\ast A$, it matters whether $A$ passes in front of or behind $B$. The isomorphisms $\gamma^{\ast}_{A,B}$ are $(\gamma^{\ast}_{B,A})\1$, known as braiding isomorphisms, are in general different. The multiplicative structure of $\A$ is \emph{braided}.

The most important aspect is related to computing.
It is not enough for us to know that there are two braiding isomorphisms from $A\ast B$ to $B\ast A$ or that there is an associativity isomorphism from $(A\ast B)\ast C$ to $A\ast(B\ast C)$. We need these isomorphisms, in matrix form with respect to the fixed bases, for computational purposes.

\subsection{Witness requirements}\mbox{}

This subsection is more specialized.
We mentioned above that the standard witnesses are subject to various requirements. One may wonder what are those requirements. We give the appropriate references.

For the additive structure, the appropriate requirements were found by Mac Lane \cite{Maclane1963} and subsequently simplified by Kelly \cite{Kelly}. For the braided multiplicative structure, the requirements were supplied by Joyal and Street \cite{JS1,JS2}.

Multiplication interacts with addition via the distributivity laws. For the case where both the additive and multiplicative structures of $\A$ are symmetric, the requirements for distributivity have been identified by Miguel Laplaza \cite{Laplaza1,Laplaza2} who was a postdoc of Mac Lane. In \cite{G238}, we identified appropriate requirements for distributivity in the case where the additive structure is symmetric but the multiplicative structure is braided.

\subsection{The additive structure}\mbox{}

This subsection is devoted to a technical issue of independent interest.

We hoped that the addition operation $+$ on $\A$ can be taken to be literally (not only up to isomorphism) commutative and associative, that is, that we can get by with the identity witnesses for the commutativity and associativity of addition. Unfortunately this is impossible.

For illustration of what goes wrong with the identity witnesses, we simplify the example described above.  Let's abstract from vector spaces and concentrate on their fixed bases: finite sets with disjoint union as addition. We recall the standard definition of disjoint union of sets.

\medskip\noindent{\tt Definition~1.}
The disjoint union of sets $A,B$ is the set\\[1ex]
\phantom{\tt Definition~1}\quad $\displaystyle A+B = \{(a,0): a\in A\} \cup \{(b,1): b\in B\}. $ \textqed

\begin{qna}
\Qn Definition~1 does not look standard to me. If fact, it looks rather arbitrary. Instead of $0$ and $1$, I can use different tags, say, $1$ and $2$.

\Ar It takes a bit of category theory to explain the standard character of the definition. For any choice of the two tags, there is a natural enhancement of the definition  with canonical embeddings of $A$ and $B$ into $A+B$; the resulting operation has the universal property of the coproduct. That is what makes the definition, in any of these variations, standard.
\end{qna}

\medskip
This disjoint union of Definition~1 is neither commutative nor associative.
One may think that there is no definition that is better in the sense that it makes disjoint union literally, not just up to isomorphism, commutative and associative. But such a ``better'' definition does exist. Let $\N$ be the set of natural numbers, i.e., nonnegative integers.

\medskip\noindent{\tt Definition~2.}
The disjoint union of finite sets $A,B$ is the set\\[1ex]
\phantom{\tt Definition~2}\quad $\displaystyle A\dot{+}B = \{n\in\N: n < |A| + |B|\}$. \textqed

\medskip
Let's adopt the set-theoretic convention that a natural number is the set of smaller natural numbers. Then Definition~2 says that the disjoint union $A\dot{+}B$ is the number $|A|+|B|$.

It is easy to see that $A\dot{+}B$ is indeed commutative and associative, so that the standard witnesses for the commutativity and associativity can be taken to be identities.
Unfortunately, we cannot push our luck too far. For example, if a finite set $A$ is not a number then the equality $A = A\dot{+}\emptyset$ cannot be witnessed by the identity. For, if $A$ is identical to $A\dot{+}\emptyset$, then $A$ is a number. Furthermore, there is no canonical embedding of $A$ into $A\dot{+}B$.

\begin{qna-}
\Qn How about generalizing sets to multisets? There is, I think, a natural disjoint union of multisets which is commutative and
associative.

\Ar Recall that our finite sets are fixed bases of vector spaces.

\Qn I see the problem. A multiset basis of a vector space does not make much sense.
\end{qna-}

\section{Summary}
\label{sec:finale}

As we mentioned above, categories were introduced by Samuel Eilenberg and Saunders Mac Lane as an auxiliary notion in their general theory of natural equivalences \cite{EM}.
Here we argue that something like categories is needed on a more basic level.

As you work with operations on structures, it may be necessary to coherently manipulate witnesses for various properties of these operations. We mentioned associativity, commutativity and distributivity, but many additional properties are in play in topological quantum computing and elsewhere. The coherent witness-manipulation problem may be hard.

This necessity of coherent witness-manipulation cannot be proven mathematically, and in some cases one can get around the coherent witness-manipulation problem. For example, for limited purposes, the narrow problem of a reasonable definition of commutative and associative disjoint union of sets can be solved by generalizing sets to multisets. Unfortunately this solution is of little help if the sets in question are vector-space bases.

In general, a working mathematician, to use Mac Lane's term \cite{Maclane1971}, is well advised to be aware of the coherent witness-manipulation problem and to know that category theory or something similar provides an appropriate framework to address the problem. Of course, the working mathematician in question may be a computer scientist or physicist.

\begin{qna-}
\Qn What do you mean by something similar to category theory?

\Ar We didn't want to rule out possible alternatives. In some situations, it suffices to consider groupoids, i.e., to restrict attention to isomorphisms. This setting can be presented in a way closer to traditional algebra \cite{G239}.

\Qn Is there an objective need to deal with more general homomorphisms?

\Ar Yes, isomorphisms are sometimes insufficient. Consider, for example, Definition~2 of disjoint union. Why does it feel so lousy? One reason is that it does not say where $A$ and $B$ are in the disjoint union. To have a useful disjoint union, one needs even more, namely where individual elements of $A$ and $B$ lie in the disjoint union. That information amounts to embeddings of $A$ and $B$ into the disjoint union, and those are not isomorphisms.
\end{qna-}

\subsection*{Acknowledgement} We thank Samson Abramsky, John Baez, Bob Coecke and Prakash Panangaden for their comments.

\end{document}